\documentclass[%
 aip,
 amsmath,amssymb,
 reprint,%
]{revtex4-1}

\usepackage{graphicx}
\usepackage{dcolumn}
\usepackage{bm}

\usepackage[utf8]{inputenc}
\usepackage[T1]{fontenc}
\usepackage{mathptmx}
\usepackage{etoolbox}

\makeatletter
\def\@email#1#2{%
 \endgroup
 \patchcmd{\titleblock@produce}
  {\frontmatter@RRAPformat}
  {\frontmatter@RRAPformat{\produce@RRAP{*#1\href{mailto:#2}{#2}}}\frontmatter@RRAPformat}
  {}{}
}%
\makeatother
\begin{document}

\preprint{AIP/123-QED}

\title[Physics-informed neural networks for imaging and parameter retrieval of photonic nanostructures from near-field data]{Physics-informed neural networks for imaging and parameter retrieval of photonic nanostructures from near-field data}
\author{Yuyao Chen}
 \affiliation{Department of Electrical \& Computer Engineering and Photonics Center, Boston University, 8 Saint Mary's Street, Boston, Massachusetts, 02215, USA}
\author{Luca Dal Negro}%
\email{dalnegro@bu.edu}
\affiliation{Department of Electrical \& Computer Engineering and Photonics Center, Boston University, 8 Saint Mary's Street, Boston, Massachusetts, 02215, USA}
\affiliation{Division of Materials Science \& Engineering, Boston University, 15 Saint Mary's St., Brookline, Massachusetts 02446, USA}%
\affiliation{Department of Physics, Boston University, 590 Commonwealth Avenue, Boston, Massachusetts, 02215, USA}

\date{\today}

\begin{abstract}
In this paper, we develop a deep learning approach for the accurate solution of challenging problems of near-field microscopy that leverages the powerful framework of physics-informed neural networks (PINNs) for the inversion of the complex optical parameters of nanostructured environments. Specifically, we show that PINNs can be flexibly designed based on the full-vector Maxwell's equations to inversely retrieve the spatial distributions of the complex electric permittivity and magnetic permeability of unknown scattering objects in the resonance regime from near-field data. Moreover, we demonstrate that PINNs achieve excellent convergence to the true material parameters under both plane wave and point source (localized) excitations, enabling parameter retrieval in scanning near-field optical microscopy (SNOM). Our method is computationally efficient compared to traditional data-driven deep learning approaches as it requires only a single dataset for training. Furthermore, we develop and successfully demonstrate adaptive PINNs with trainable loss weights that largely improve the accuracy of the inverse reconstruction for high-index materials compared to standard PINNs. Finally, we demonstrate the full potential of our approach by retrieving the space-dependent permittivity of a three-dimensional (3D) unknown object from near-field data. The presented framework paves the way to the development of a computationally-driven, accurate, and non-invasive platform for the simultaneous retrieval of the electric and magnetic parameters of resonant nanostructures from measured optical images, with applications to biomedical imaging, optical remote sensing, and characterization of metamaterial devices.
\end{abstract}

\maketitle

In the past decades, the engineering of electromagnetic waves in optical materials heavily relied on either analytical theories or numerical methods to obtain the solutions of physics-based partial differential equations (PDEs) models. However, it has become increasingly difficult to apply these traditional approaches to complex optical structures and heterogeneous media, particularly in relation to the challenging inverse problems of near-field optical microscopy with many applications to biomedical imaging, material characterization, and nano-optical device inspection \cite{courjon2003near,klein2017dual,stanciu2020characterization,zhang2020visibility}. Specifically, the parameter retrieval problem of near-field microscopy consists in estimating the properties of a scattering object, usually identified by its shape and dielectric permittivity, from a limited set of near-field data under different excitation conditions. Thanks to the impressive developments of the SNOM technique \cite{novotny2012principles}, it is currently possible to image the amplitude and phase of the scattered fields of complex nanostructures in the near-field zone with nanoscale spatial accuracy \cite{ASH1972,betzig1992near,dallapiccola2009near,novotny2012principles,Govyadinov2014,Klein2014,klein2017dual,stanciu2020characterization,zhang2020visibility}. In particular, by exciting and detecting photonic structures locally, near-field techniques  provide a ``non-invasive" approach for the characterization of complex nanostructures. However, due to the strong multiple scattering of radiation in multi-particle systems at large refractive index contrast, the problem of retrieving the material optical constants from near-field data becomes intrinsically non-linear and ill-posed. As a result, iterative optimization methods are proposed to solve these inverse problems with good accuracy, but they are usually computationally expensive \cite{kamilov2015learning,molesky2018inverse,liu2017seagle, kamilov2017plug, Pham2018Versatile, colton2018looking}. On the other hand, there is a growing interest in developing deep learning (DL) algorithms for electromagnetic wave engineering. This rapidly emerging approach includes training artificial neural networks (ANNs) to solve the photonics inverse problems \cite{sun2018efficient, sanghvi2019embedding,lim2020three,mastel2018understanding,Ma2021,Liu2021Tackling}. Although demonstrated successful at solving several inverse problems \cite{Liu2021Tackling,Ma2021}, DL methods are essentially data-driven techniques that require a time-consuming training process in order to instruct ANNs using massive datasets. In order to improve on purely data-driven DL methods, it is important to constrain them by leveraging the underlying physics of the problems, thus relaxing the burden on the training and data acquisition steps. Therefore, it is critical to build a robust framework that efficiently integrates powerful ANN architectures with the physical laws that fundamentally constrain the complex parameter retrieval problems of near-field microscopy.
In this context, physics-informed neural networks (PINNs) is a general framework developed for solving both forward and inverse problems that are mathematically modeled by arbitrary PDEs of integer or fractional orders \cite{Raissi,lu2021deepxde,Pang2019fPINNs}. In our previous work we have applied PINNs to retrieve purely real permittivities of lossless materials from real-valued electric field observations under plane wave excitation \cite{chen2020physics}. 

In this paper, we develop a more general PINNs framework for solving the parameter retrieval problems of near-field optical microscopy that are of direct interest to biomedical imaging and nanotechnology. In particular, we address and demonstrate the accurate retrieval of the complex electric permittivity $\varepsilon_{r}$ of resonant nanostructures based on complex wave equations using complex electric field (synthetic) data obtained from forward finite element method (FEM) simulations. Furthermore, we demonstrate the retrieval of the complex magnetic permeability $\mu_{r}$ of resonant nanostructures based on the full inversion of Maxwell's equations. We show that PINNs retrieve correctly the material parameters from near-field data sampled under both plane wave and localized (line current source) excitations, which can enable inverse parameter retrieval using the dual-SNOM technique \cite{dallapiccola2009near,Klein2014,klein2017dual}. Importantly, we show the simultaneous inverse retrieval of the space-dependent materials parameters $\varepsilon_{r}$ and $\mu_{r}$ without prior shape information in a two-dimensional (2D) geometry. In this context, an adaptive PINNs algorithm is proposed and developed to improve the stability and accuracy in retrieving high-index material parameters in a regime where the standard PINNs method fails. Finally, we demonstrate the full potential of our method by  successfully retrieving the complex permittivity profile of an unknown 3D scattering object from sampled synthetic data. We remark that the framework shown here is more general than the one shown in our previous work \cite{chen2020physics} and can be used to simultaneously retrieve the space-dependent complex optical parameters of unknown objects with electric and magnetic resonant responses in the presence of losses and under different excitation conditions. All the implementations of the developed PINN algorithms used in this paper are obtained within the powerful DeepXDE library \cite{lu2021deepxde}. 


\begin{figure}[t]
\centering
\includegraphics[width=\linewidth]{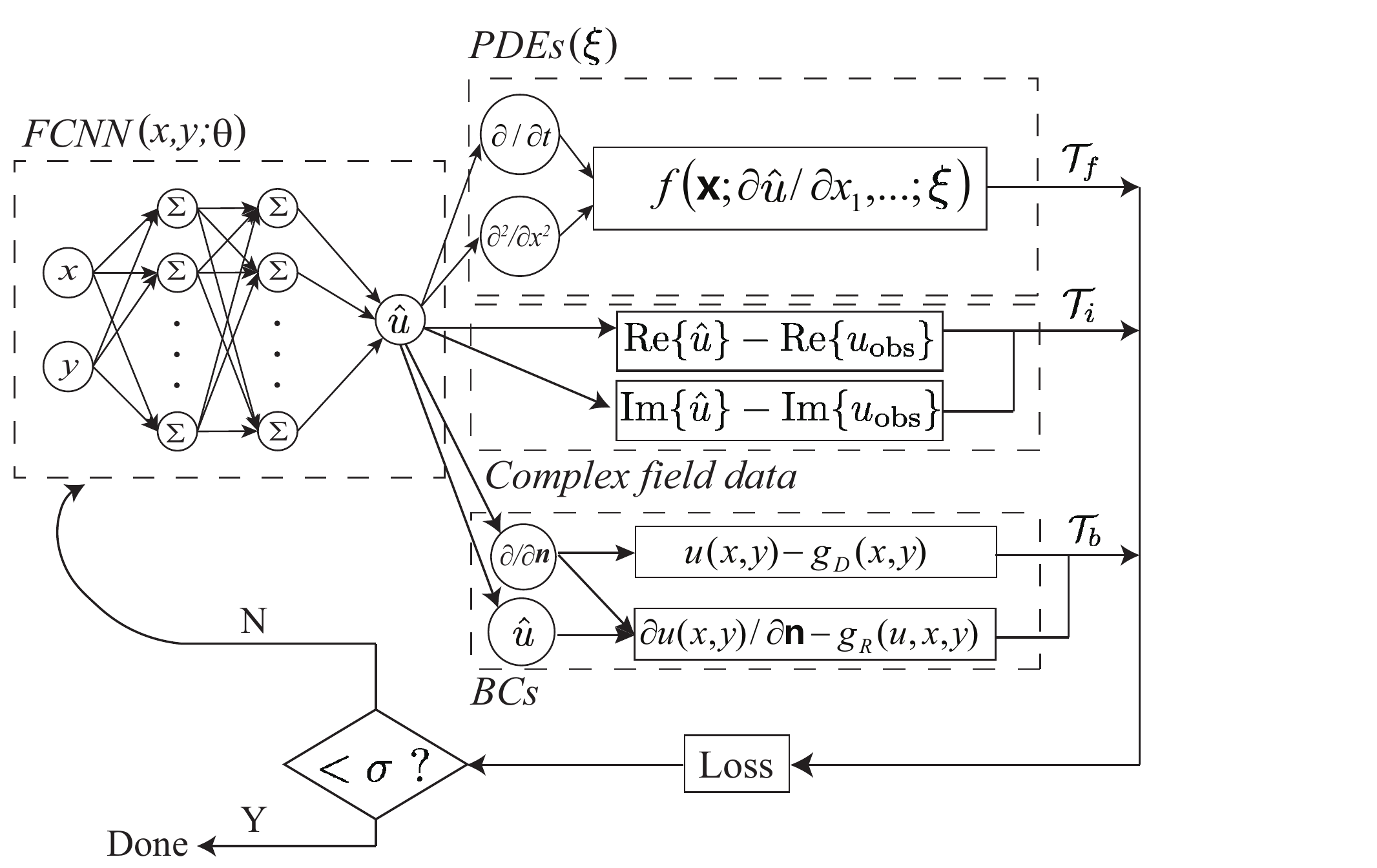}
\caption{(a) schematics of PINN solving the parameter retrieval problem for near-field microscopy. The FCNN$(x,y;\theta)$ denotes the fully connected neural network with its output as the PDE surrogate solution for the inverse problem. The 'Loss' symbol in the bottom represents the loss function that restricts $\hat{u}$ to satisfy the PDEs, complex field data, and boundary conditions (BCs). During the training process, both $\mathbf{\theta}$ and the unknown object material parameter $\varepsilon_r$ in the neural networks are optimized until the value of loss function is below certain threshold $\sigma$.}
\label{Fig1}
\end{figure}

The general PINN algorithm utilized for solving inverse PDE problems is schematically illustrated in Fig. \ref{Fig1}. We first construct an artificial neural network (ANN) with input $\mathbf{x}=(x_1,y_1;x_2,y_2;\dots;x_N,y_N)$ that are either grid points or random points sampled on the studied domain $\Omega$ and output $\hat{u}(\mathbf{x};\boldsymbol{\theta},\xi)$ as a surrogate of the PDE solution $u(\mathbf{x})$ \cite{lu2021deepxde}. Here $\boldsymbol{\theta}$ denotes a vector containing all the weights and biases in the ANN and $\xi$ describes the unknown parameters in the PDEs that need to be retrieved. A simple fully connected neural network (FCNN) is employed here but the method can conveniently be extended to accommodate more complex ANN architectures. Crucially, we have built the loss function that constrains  $\hat{u}$ to satisfy the PDEs describing the electromagnetic physics of the considered problems. 

As a first example, we consider the near-field microscopy parameter retrieval problem of retrieving the complex relative permittivity $\varepsilon_{r}$ defined on a domain $\Omega\subset \mathbf{R}^2$ with the constraint:
\begin{equation}\label{eq:pde}
f\left( \mathbf{x}; \frac{\partial \hat{u}}{\partial x}, \frac{\partial \hat{u}}{\partial y}, \frac{\partial^2 \hat{u}}{\partial x^2}, \frac{\partial^2 \hat{u}}{\partial y^2}; \varepsilon_{r} \right) = 0, \quad \mathbf{x} \in \Omega.
\end{equation}
with boundary conditions $\mathcal{B}(u, \mathbf{x})=0$ on $\partial \Omega$, where $\varepsilon_r$ corresponds to the $\xi$ parameter in PINNs and the function $f$ is derived from wave equation as it will be discussed later sections. The derivatives of $\hat{u}(\mathbf{x};\boldsymbol{\theta},\varepsilon_{r})$ in Eq. \ref{eq:pde} are obtained using the auto differentiation of the ANN, which is already implemented in the TensorFlow package \cite{Abadi}. We define the loss function that constraints our PINNs formulations by:
\begin{eqnarray}\label{eq:loss}
\mathcal{L}(\boldsymbol{\theta},w_f,w_i,w_b, \varepsilon_{r}) = & w_f \mathcal{L}_f(\boldsymbol{\theta},\varepsilon_{r}; \mathcal{T}_f) + w_i \mathcal{L}_i(\boldsymbol{\theta},\varepsilon_{r}; \mathcal{T}_i) \nonumber\\ &  + w_b \mathcal{L}_b(\boldsymbol{\theta},\varepsilon_{r}; \mathcal{T}_b),
\end{eqnarray}
where $w_f$, $w_b$, and $w_i$ are the loss weights and 
\begin{eqnarray}
\mathcal{L}_f(\boldsymbol{\theta},\varepsilon_{r}; \mathcal{T}_f) &= \frac{1}{|\mathcal{T}_f|} \sum_{\mathbf{x} \in \mathcal{T}_f} \left\| f\left( \mathbf{x}; \frac{\partial \hat{u}}{\partial x}, \frac{\partial \hat{u}}{\partial y}, \frac{\partial^2 \hat{u}}{\partial x^2}, \frac{\partial^2 \hat{u}}{\partial y^2}; \varepsilon_{r} \right)  \right\|^2
\end{eqnarray}
\begin{eqnarray}
\mathcal{L}_{i}(\boldsymbol{\theta},{\varepsilon_{r}}; \mathcal{T}_i) &= \frac{1}{|\mathcal{T}_i|} \sum_{\mathbf{x} \in \mathcal{T}_i} \| Re\{\hat{u}(\mathbf{x})\} - Re\{u_{obs}(\mathbf{x})\} \|^2 \nonumber\\ 
&+\| Im\{\hat{u}(\mathbf{x})\} - Im\{u_{obs}(\mathbf{x})\} \|^2, \label{complex field observations}
\end{eqnarray}
\begin{eqnarray}
\mathcal{L}_{b}\left(\boldsymbol{\theta},\varepsilon_{r}; \mathcal{T}_{b}\right)&=\frac{1}{\left|\mathcal{T}_{b}\right|} \sum_{\mathbf{x} \in \mathcal{T}_{b}}\|\mathcal{B}(\hat{u}, \mathbf{x})\|_{2}^{2}.
\end{eqnarray}
$\mathcal{L}_f$, $\mathcal{L}_i$, and $\mathcal{L}_b$ denote the $L^2$ norm of residuals for the PDEs, the complex field observations, and the BCs, respectively. The symbols $Re\{\cdot\}$ and $Im\{\cdot\}$ denote the real and imaginary parts of a complex quantity, respectively. We can obtain the complex field observations $u_{obs}(\mathbf{x})$ from experimental or numerical simulations (i.e., synthetic data). The quantities $\mathcal{T}_{f}$, $\mathcal{T}_{i}$, $\mathcal{T}_{b}$ denote the residual points for $\mathcal{L}_f$, $\mathcal{L}_i$, and $\mathcal{L}_b$, respectively \cite{lu2021deepxde}. In the last step, we train the neural networks of PINNs to search for the parameters $\mathbf{\theta}$ and $\varepsilon_{r}$ that minimize the total loss function specified in Eq. \ref{eq:loss}. As we will show in the following sections, the proposed framework solves the parameter retrieval problem of near-field microscopy using only one set of complex field observations. We remark that using PINNs we can solve a complex inverse problem at at a small computational cost compared to the solution of the associated forward one, since the only difference between the two is the introduction of the extra loss term $\mathcal{L}_{i}$ in the Eq. \ref{eq:loss}. 

\section{Results and discussion}
\subsection{Objects with known shapes}
\subsubsection{Retrieval of the complex electric permittivity}\label{section: epsilon single}
\begin{figure}[t!]
\centering
\includegraphics[width=\linewidth]{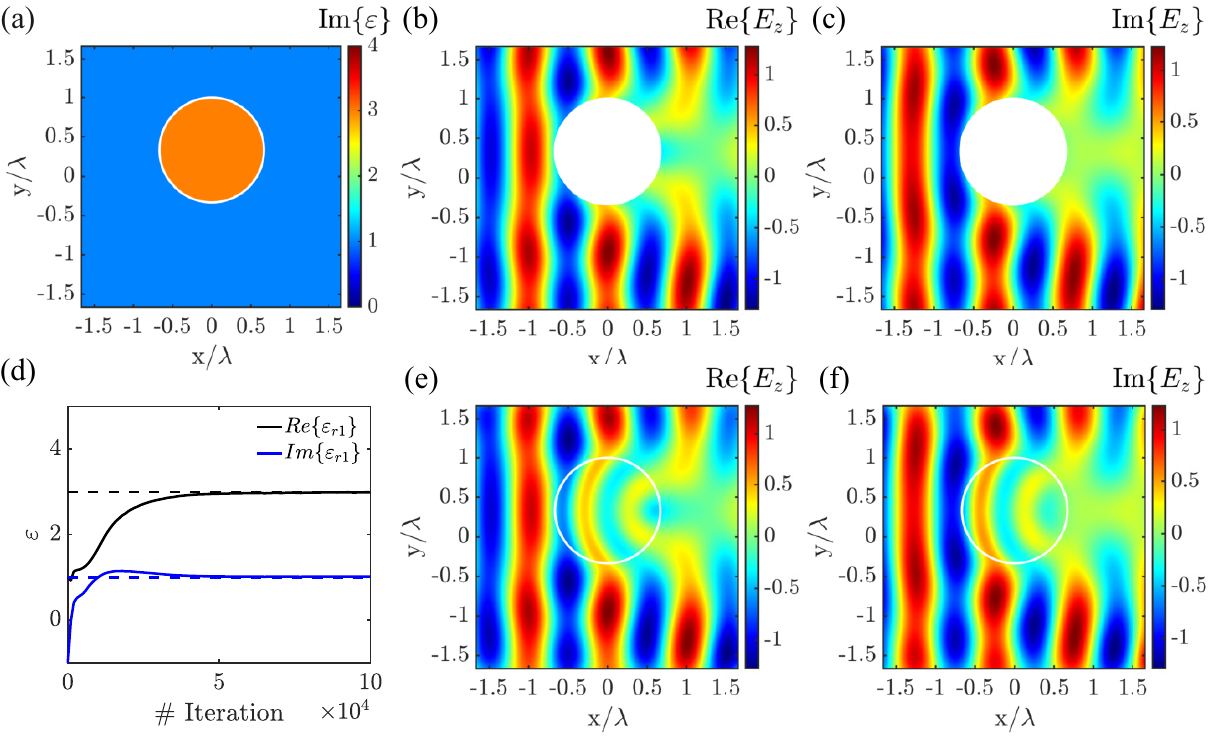}
\caption{(a) real part of the $\varepsilon_r$ profile used for the FEM forward scattering simulation. (b,c) real and imaginary parts of the complex electric field $E_z$ respectively, used to train PINNs. The blank regions in (b) and (c) indicate that the $E_z$ data inside the cylinder are excluded. (d) inverse retrieval of the complex dielectric function with respect to the number of iterations. (e,f) real and imaginary parts of the complex $E_z$ field reconstructed by PINNs after $10^4$ iterations.
}
\label{Fig2}
\end{figure}

We recently utilized the powerful PINN method for solving the inverse Mie scattering problem limited to lossless materials \cite{chen2020physics}. Here we extend the approach by taking into account the losses of the materials, which require the more difficult inversion of complex optical parameters. We start by considering the case of retrieving the complex permittivity of a single dielectric cylinder with a diameter comparable with the wavelength of light. Specifically, we study a cylinder with radius $r=2{\mu}m$ and $\varepsilon_r = 3 + 1j$ under TM polarized plane wave excitation at wavelength $\lambda = 3{\mu}m$. We obtain synthetic data by performing forward simulations using the FEM modeling \cite{chen2020physics} (see the Methods section for additional details on the FEM simulations). The $\varepsilon_r$ real part profile of the simulated structure is shown in Fig. \ref{Fig2} (a). We denote the region of vacuum ($Re\{\varepsilon_r\}=1$) as $\Omega_1$ and the region occupied by the cylinder ($Re\{\varepsilon_r\}=3$) as $\Omega_2$. 

The dynamic Maxwell's equations allow one to derive the wave equation for a non-homogeneous medium in the form \cite{johndavidjackson1998}:
\begin{equation}\label{nonhomogeneous 2D}
\nabla^{2} \mathbf{E}-\mu \epsilon \frac{\partial^{2} \mathbf{E}}{\partial t^{2}}=-\nabla( \mathbf{E}\cdot\nabla {\ln\epsilon})-\mu \frac{\partial \mathbf{H}}{\partial t} \times \nabla \ln \mu
\end{equation}
where $\mathbf{E}$ is the electric field, $\mathbf{H}$ is the magnetic field, and $\varepsilon(\mathbf{r})=\varepsilon_0\varepsilon_r(\mathbf{r})$ and $\mu(\mathbf{r})=\mu_0\mu_r(\mathbf{r})$ are the space-dependent medium permittivity and permeability, respectively. The symbols $\mu_0$ and $\varepsilon_{0}$ denote the permeability and permittivity of free space, respectively. Without loss of generality, we first study the parameter retrieval in 2D geometries for retrieving the complex parameter $\varepsilon_r$ under the TM polarization excitation $\mathbf{E}(\mathbf{r})=E_z(x,y)$, which yields $\mathbf{E}\cdot\nabla\ln\varepsilon(x,y)=0$. 
Besides, we assume at first that we are studying the a non-magnetic object with relative permeability $\mu_r(\mathbf{r})=1$ for which $\nabla \ln \mu=0$, yielding the wave equation:
\begin{equation}\label{wave_equation}
\nabla^{2} E_z- \mu\varepsilon\frac{\partial^{2} E_z}{\partial t^{2}}=0
\end{equation}
Separating the real and imaginary parts of the wave equation, we obtain the following PDE model that we have implemented in our PINNs:
\begin{align}
    \nabla^{2}Re\{E_z\}+\left[Re\{E_z\}Re\{\varepsilon_{r}\}-Im\{E_z\}Im\{\varepsilon_{r}\}\right]k_{0}^{2}&=0 \label{wave eqn real}\\
    \nabla^{2}Im\{E_z\}+\left[Im\{E_z\}Re\{\varepsilon_{r}\}+Re\{E_z\}Im\{\varepsilon_{r}\}\right]k_{0}^{2}&=0 \label{wave eqn imag}
\end{align} 
where $k_0=\frac{2\pi}{\lambda}$ is the incident wave number. Since the shape of the object is known a priori, we denote by $\varepsilon_{r1}$ and $\varepsilon_{r2}$ the complex homogeneous permittivity in regions $\Omega_1$ and $\Omega_2$, respectively, and we impose the electromagnetic boundary conditions (BCs) at the boundary $\partial\Omega_2$ as follows:
\begin{align} 
\left.Re\{E_{z}^{(1)}\}\right|_{\mathbf{x}=\partial\Omega_2}&=\left.Re\{E_{z}^{(2)}\}\right|_{\mathbf{x}=\partial\Omega_2}\label{wave eqn BCs:E real}\\ \left.Im\{E_{z}^{(1)\}}\right|_{\mathbf{x}=\partial\Omega_2}&=\left.Im\{E_{z}^{(2)}\}\right|_{\mathbf{x}=\partial\Omega_2} \label{wave eqn BCs:E imag}\\
\left.\frac{\partial Re\{E_{z}^{(1)}\}}{\partial r}\right|_{\mathbf{x}=\partial\Omega_2}&=\left.\frac{\partial Re\{E_{z}^{(2)}\}}{\partial r}\right|_{\mathbf{x}=\partial\Omega_2}\label{wave eqn BCs:H real}\\ 
\left.\frac{\partial Im\{E_{z}^{(1)}\}}{\partial r}\right|_{\mathbf{x}=\partial\Omega_2}&=\left.\frac{\partial Im\{E_{z}^{(2)}\}}{\partial r}\right|_{\mathbf{x}=\partial\Omega_2} \label{wave eqn BCs:H imag}
\end{align} 
where $E_z^{(k)},~(k=1,2)$ are the complex electric fields in domain $\Omega_k$ and $r$ is the radial component in polar coordinates with its origin at the  center of the cylinder. 
The real and imaginary parts of $E_z$ obtained from FEM simulations and utilized for training PINN are displayed in Fig. \ref{Fig2} (b) and (c), respectively. We sample the complex $E_z$ on a square grid in the $\Omega_1$ region with resolution $\Delta x=0.02 \lambda$ as the training dataset, which is achievable using current near-field microscopy techniques \cite{mastel2018understanding,Ma2021}. We set the initial value $\varepsilon_{r2}=1-j1$ and train the FCNN by minimizing the loss function specified in Eq. \ref{eq:loss}. Further details on the training methods and the utilized hyperparameters of the FCNN are described in the Methods section. The retrieved values of the complex permittivity $\varepsilon_{r2}$ with respect to the iteration number are shown in Fig. \ref{Fig2} (d) where the values of the true solution are indicated by the dashed lines. The results displayed in Fig. \ref{Fig2} demonstrate the accurate retrieval of the complex permittivity of non-ideal lossy dielectric materials using PINNs. Moreover, the developed PINN model successfully reconstructed the complex $E_z$ distribution within $\Omega_2$ after training, as we show in Figs. \ref{Fig2} (e) and (f) for the real and imaginary parts of $E_z$, respectively. The $L^2$ errors of the $E_z$ between data obtained from FEM simulations and the predictions of PINNs are $\sim 10^{-4}$ (The $L^2$ error definition and exact values for $E_z$ real and imaginary parts are discussed in supplementary material). Therefore, we conclude that PINNs can be reliably extended to obtain the electric permittivity of resonant nanostructures from complex electric field data. In the next section, we will introduce an even more general PINN model for the simultaneous retrieval of the complex optical parameters of electric and magnetic scattering media.

\subsubsection{Simultaneous retrieval of permittivity and permeability}\label{section: magnatic material single parameters}

In this section we consider the retrieval of both the $\mu_r$ and $\varepsilon_r$ of magnetic materials that have important applications to biomedical, environment treatment, and nanotechnology \cite{Li2016Current,MOHAMMED20171,Cardoso2018Advances,MAKSYMOV201636}. Since the parameters $\mu_r$ and $\varepsilon_r$ are coupled in the Eq. \ref{wave_equation}, we now consider a PDE model based on the dynamic Maxwell's equations, where $\mu_r$ and $\varepsilon_r$ appear separately. In particular, for the TM polarization the PDEs are as follows:
\begin{align}
\frac{\partial E_{z}}{\partial y}&=j \omega \mu_{0} \mu_{rk} H_{x} \label{Maxwell 2D:1}\\
-\frac{\partial E_{z}}{\partial x}&=j \omega \mu_{0} \mu_{rk} H_{y} \label{Maxwell 2D:2}\\
\frac{\partial H_{y}}{\partial x}-\frac{\partial H_{x}}{\partial y}&=-j \omega \varepsilon_{0} \varepsilon_{rk} E_{z} \label{Maxwell 2D:3}
\end{align}
We denote the relative permeability and permittivity of region $\Omega_k$ as $\mu_{rk}$ and $\varepsilon_{rk}~(k=1,2)$, respectively. Here $H_x$ and $H_y$ represent the $x$ and $y$ components of the magnetic field, respectively. We simulate a cylinder with the same geometry as shown in Fig. \ref{Fig2} (a) and we set $\varepsilon_{r1}=\mu_{r1}=1$, $\varepsilon_{r2}=1$, and $\mu_{r2}=2.5+0.6j$. The same notations of $\Omega_1$ for the vacuum region and $\Omega_2$ for the cylinder region are used. We can directly apply the following BCs at the cylinder boundary $\Omega_2$ to constrain the PINN's retrieval of the optical constants using:
\begin{eqnarray} 
\left.Re\{E_{z}^{(1)}\}\right|_{\mathbf{x}=\partial\Omega_2}&=\left.Re\{E_{z}^{(2)}\}\right|_{\mathbf{x}=\partial\Omega_2}\\
\left.Im\{E_{z}^{(1)\}}\right|_{\mathbf{x}=\partial\Omega_2}&=\left.Im\{E_{z}^{(2)}\}\right|_{\mathbf{x}=\partial\Omega_2} \label{maxwell eqn BCs:E}\\
\left.\overrightarrow{n}\times Re\{\mathbf{H}^{(1)}\}\right|_{\mathbf{x}=\partial\Omega_2}&=\left.\overrightarrow{n}\times Re\{\mathbf{H}^{(2)}\}\right|_{\mathbf{x}=\partial\Omega_2}\\ \left.\overrightarrow{n}\times Im\{\mathbf{H}^{(1)}\}\right|_{\mathbf{x}=\partial\Omega_2}&=\left.\overrightarrow{n}\times Im\{\mathbf{H}^{(2)}\}\right|_{\mathbf{x}=\partial\Omega_2} \label{wave eqn BCs:H}
\end{eqnarray} 
where $\overrightarrow{n}$ is the unit vector normal to $\partial\Omega_2$ and $\mathbf{H}^{k}=(H_x^{(k)},H_y^{(k)}),~(k=1,2)$ is the magnetic field vector in the region $\Omega_k$. 
\begin{figure}[h!]
\centering
\includegraphics[width=\linewidth]{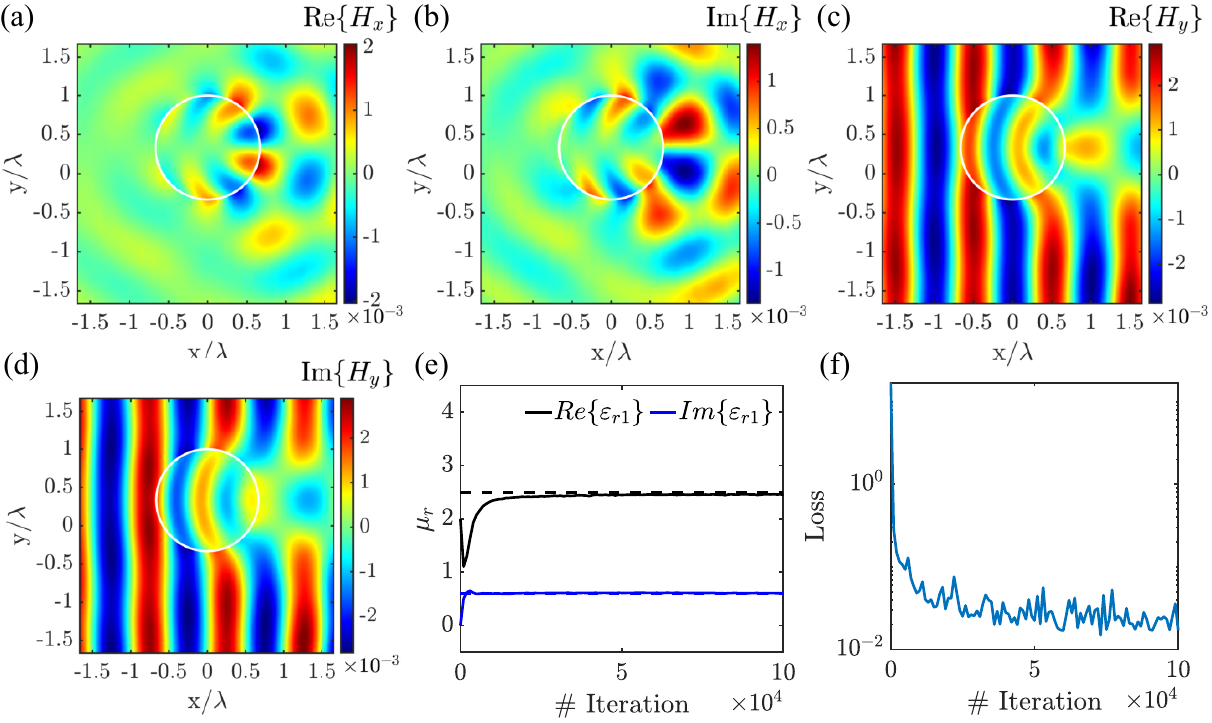}
\caption{(a,b) real and imaginary parts of the complex magnetic field $H_x$ component used in PINNs, respectively. (c,d) real and imaginary parts of the complex magnetic field $H_y$ component used in PINNs, respectively. (e) retrieval of the complex permittivity $\varepsilon_{r2}$ and permeability $\mu_{r2}$ with respect to the number of iterations. The dashed lines indicate the true values of retrieved parameters. (f) total loss value during the training process with respect to the iteration number.}
\label{Fig4}
\end{figure}

We perform the FEM simulations with details specified in the Methods section and obtain the complex $E_z$, $H_x$, and $H_y$ field data for training the network. Similar to the approach of retrieving $\varepsilon_r$, we sampled these fields only in the region $\Omega_2$ considered as the training dataset. We train the FCNN (see the Methods section for more details) to retrieve the complex $\mu_{r2}$ simultaneously by fixing $\varepsilon_{r1}=\mu_{r1}=1$ and $\varepsilon_{r2}=1$. The $\mu_{r2}$ starts from the initial value $\mu_{r2}=2$. The reconstructed complex $E_z$ profile obtained from the PINNs are shown in the supplementary material. We display the reconstructed complex $H_x$ and $H_y$ field profiles in Fig. \ref{Fig4} (a-b) and (c-d), respectively. We show the parameters retrieval during the training process in Fig. \ref{Fig4} (e), where very good convergence to the complex $\mu_{r2}$ true solutions (indicated by the dashed lines) is obtained. We quantify the convergence by computing the maximum $L^2$ error norm between the reconstructed fields from PINNs and FEM simulations, which we found to be $8 \times 10^{-3}$ (Detailed values are reported in the supplementary material). The scaling of the total loss with respect to the iteration numbers are displayed in Fig. \ref{Fig4} (f) where a satisfactory loss of $\approx{10^{-2}}$ is reached. Therefore, we have shown that the proposed framework can be implemented to solve the near-field parameter retrieval problem for both the electric and magnetic properties of an object of a given shape based on the Maxwell's equations under the plane wave excitation. In the next section, we will solve such problem under a local source excitation that is directly relevant for the inverse parameter retrieval using the SNOM technique.

\subsubsection{Permittivity retrieval using dual-SNOM setup}\label{section: line source excitation}
In the previous section we  discussed the parameter retrieval of complex $\varepsilon_r$ and $\mu_r$ from the near-field data under plane wave excitation. However, in many applications of the SNOM technique, complex nanostructures are often illuminated using an excitation probe in the near-field \cite{ASH1972,betzig1992near,dallapiccola2009near,novotny2012principles,Klein2014,klein2017dual,stanciu2020characterization,zhang2020visibility,Govyadinov2014}. In particular, we address here the parameter retrieval problem of the dual-SNOM technique \cite{dallapiccola2009near,Klein2014,klein2017dual}, that uses two localized probes for the simultaneous excitation and detection of the investigated nanostructures. 
\begin{figure}[t!]
\centering
\includegraphics[width=\linewidth]{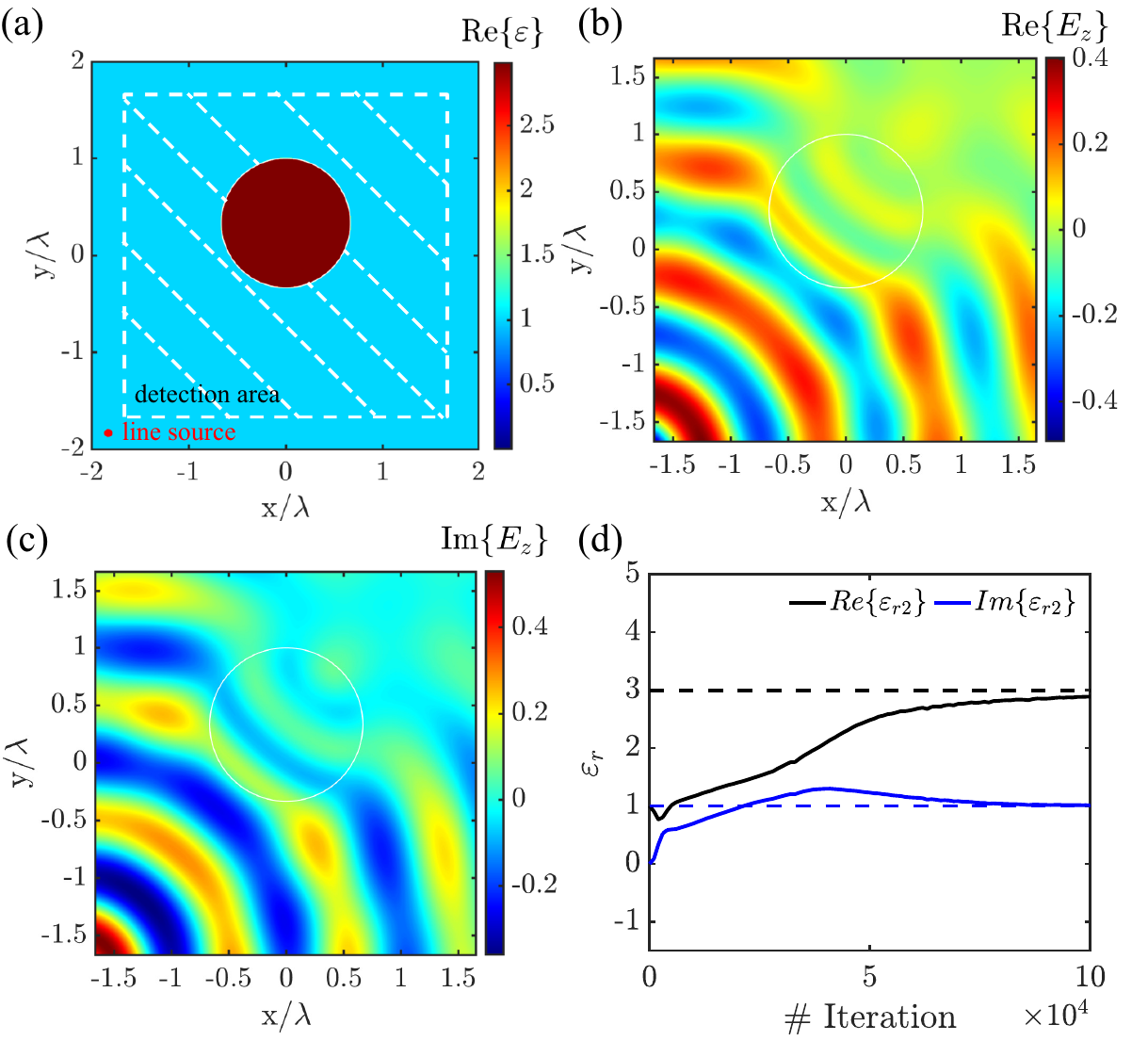}
\caption{(a) schematic of FEM simulation for the dual-SNOM setup. (b,c) real and imaginary parts of the complex electric field $E_z$ in the dectection area obtained from PINNs, respectively. (d) retrieval of the complex permittivity $\varepsilon_{r2}$ with respect to the number of iterations. The dashed lines indicate the true values of retrieved parameters.}
\label{FigSNOM}
\end{figure}
The FEM setup for simulating the parameter retrieval using the dual-SNOM is schematically illustrated in Fig. \ref{FigSNOM} (a). We illuminate the same photonic structure as in section \ref{section: epsilon single} with a line current source along the z-axis (out-of-plane) at the position indicated by the small red dot that emulates the excitation probe. The near-field information for training PINN is then collected by the detecting probe that is scanned across the detection area (white dashed square, excluding the cylinder domain). We keep the same field sampling resolution as in the previous examples. The wave equation models in Eqs. \ref{wave eqn real} and \ref{wave eqn imag} can still be implemented because the electric field is still TM polarized $\mathbf{E}(\mathbf{r})=E_z(x,y)$. Furthermore, the BCs described by Eqs. \ref{wave eqn BCs:E real}-\ref{wave eqn BCs:H imag} can be also applied. We train an FCNN with the same hyperparameters and training method used in the section \ref{section: epsilon single}. The obtained complex electric field $E_z$ profiles from PINN are shown in Fig. \ref{FigSNOM} (b) and (c), where we retrieve the $E_z$ profiles also inside the cylinder region (based uniquely on external field data). We evaluate that the $L^2$ errors for the $E_z$ complex field between the FEM simulation and PINNs to be $\sim 1 \times 10^{-4}$ (detailed values are given in the supplementary material). The complex $\varepsilon_r$ retrieval with respect to the number of iterations is shown in Fig. \ref{FigSNOM} (d), demonstrating the rapid convergence to the true values using the dual-SNOM configuration. The PINNs based on the wave equation can hence succesfully retrieve the $\varepsilon_r$ using only the external near-field data under different excitation conditions. Furthermore, the simultaneous retrieval of $\varepsilon_r$ and $\mu_r$ for photonic structures can also be achieved following the same approach. Therefore, the PINNs framework can be directly applied to solve the electric and magnetic material parameter retrieval problems under the dual-SNOM setup. In the next section we will address the PINNs approach to the complex parameter retrieval of resonant dielectric objects of unknown shapes.

\subsection{Objects with unknown shapes}\label{section: epsilon profile}
\subsubsection{Retrieval of complex permittivity profile}
\begin{figure}[t!]
\centering
\includegraphics[width=\linewidth]{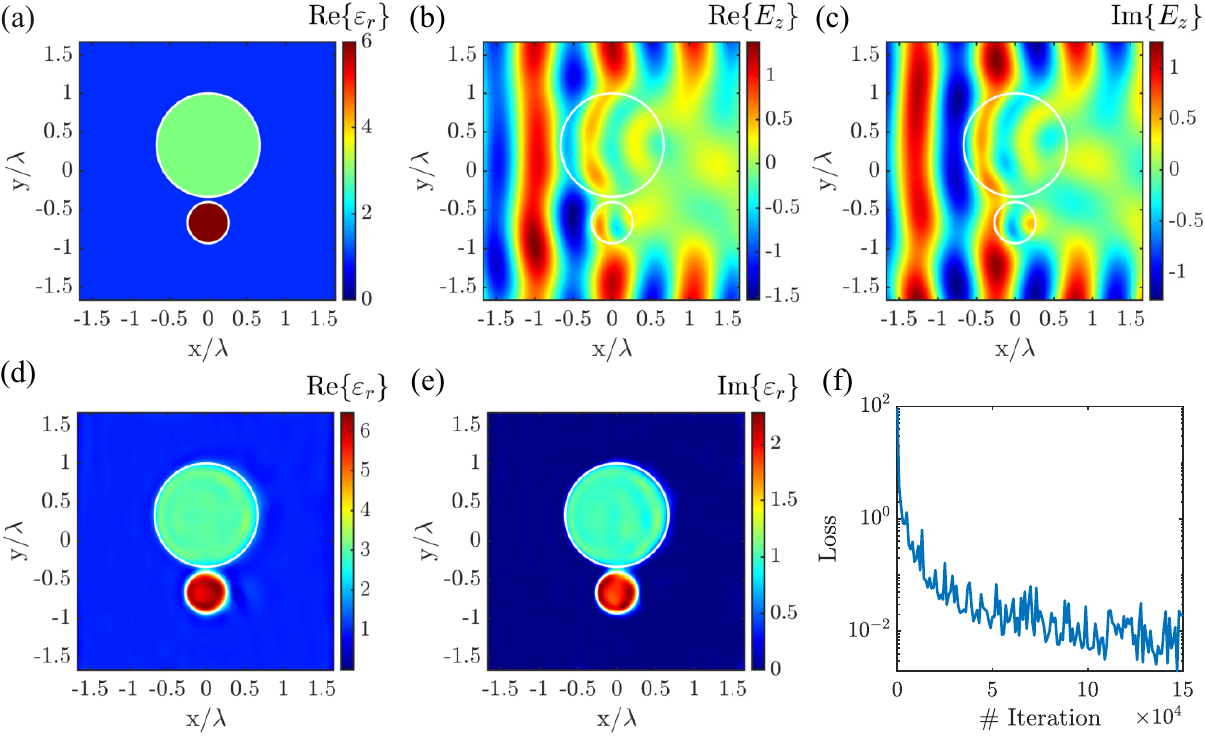}
\caption{(a) real part of the $\varepsilon_r$ profile for the two asymmetric dielectric dimer used for the FEM forward scattering simulation. (b,c) real and imaginary parts of the complex electric field $E_z$ obtained from FEM simulation. The wavelength of the incident plane wave is $\lambda=3{\mu}m$. (d,e) real and imaginary parts of the retrieved permittivity profile $\varepsilon_r$ by PINNs after $15*10^4$ iterations. (f) total loss value during the training process with respect to the iteration number in PINNs.} 
\label{Fig3}
\end{figure}

In this section we address the challenging problem of retrieving not only the complex permittivity values in the resonant regime but also the unknown shapes of the scattering objects. In particular, we will show that the proposed PINNs framework can be developed to retrieve the complex $\varepsilon_r(x,y)$ profiles of unknown objects with dimensions comparable to or larger than the wavelength of the incident light by inverting the wave equations without imposing any BCs. The efficient solution of the inverse scattering problems in non-homogeneous media shown here simultaneously addresses both the parameter retrieval and the imaging problem of microscopy that are computationally prohibitive using the traditional retrieval methods \cite{Nonlinear2006Panasyuk,Gang2007INVERSE}. In order to demonstrate this capability using PINNs we consider as a representative example the permittivity profile corresponding to the dimer configuration shown in Fig. \ref{Fig3} (a). We set the larger cylinder with radius $r_1=2{\mu}m$ and permittivity $\varepsilon_{r1}=3+j1$ centered at $(x_{c1},y_{c1})=(0,1{\mu}m)$ and the smaller cylinder radius $r_2=0.8{\mu}m$ and permittivity $\varepsilon_{r2}=6+j2$ centered at $(x_{c2},y_{c2})=(0,-2{\mu}m)$. The complex field data $E_z$ obtained from FEM simulation under TM plane wave excitation with $\lambda=3{\mu}m$ are shown in Fig. \ref{Fig3} (b) and (c) for the real and imaginary parts, respectively. More details on the FEM simulations are provided in the Methods section. 

We implement the PDE model described by Eq. \ref{wave eqn real} and Eq. \ref{wave eqn imag} with a space-dependent complex relative permittivity $\varepsilon_r\left(x,y\right)$ over the entire square domain $\Omega$. Since the shape of the object is not known \textit{a priori}, no BCs can be implemented in this case and the problem directly deals with the inversion of an non-homogeneous extended medium. We then train PINNs with only the PDE and field observation constraints and retrieve the complex $\varepsilon_r\left(x,y\right)$ profile after the training process.
To retrieve the $\varepsilon_r\left(x,y\right)$ shown in Fig. \ref{Fig3} (a), we employ the same FCNN architecture as in the last example and train it using the Adam optimizer for $1.5\times{10^{4}}$ iterations until the total loss drops below $10^{-2}$. The retrieved $\varepsilon_r$ profile real and imaginary part after training is shown in Fig. \ref{Fig3} (d) and (e), respectively. We observe that the proposed framework retrieved successfully the shape information of the dimer setup in Fig. \ref{Fig3} (d). Furthermore, we characterize the accuracy of the retrieved profiles by evaluating the complex $\varepsilon_r$ inside each cylinder domain and we obtain $\varepsilon_{r1}=(2.92\pm0.27)+j(0.96\pm0.12)$ and $\varepsilon_{r2}=(5.97\pm0.34)+j(2\pm0.15)$, which are in very good agreement with the input data. The $\varepsilon_r$ errors estimated here are the standard deviations of the corresponding quantities within each cylinder domain. We also evaluate the $L^2$ error between the $E_z$ obtained from PINNs and from the FEM simulations, which is $\sim 10\times 10^{-4}$ (see supplementary material for further details).  We display the total loss with respect to the iteration in Fig. \ref{Fig3} (f). The rapid spikes displayed by the total loss curve during the training process visibly demonstrate the highly non-linear nature of the parameter retrieval problem for near-field microscopy. Notice that we successfully retrieved the space profile of the complex permittivity $\varepsilon_r$ at almost no additional computational cost compared to the previous example shown in section \ref{section: epsilon single}, except that here we used the complex $E_z$ data over the entire $\Omega$ domain as our dataset. Therefore, the developed PINNs inversion models demonstrate accurate and efficient retrieval of both the complex permittivity values and the space distributions (i.e, shape information) of scattering objects. This achievement naturally augments near-field microscopy techniques by providing a robust, computationally driven platform for solving the imaging and the parameter retrieval problem of  dielectric structures simultaneously.

\subsubsection{Simultaneous retrieval of permittivity and permeability profiles}\label{section: magnetic material profile}
In this section we demonstrate how to improve the previous PINN setup in order to retrieve simultaneously both the $\varepsilon_r(x,y)$ and $\mu_r(x,y)$ spatial profiles, providing both electric and magnetic optical parameters together with shape information for applications to inverse near-field microscopy. In this case we must implement Eqs. \ref{Maxwell 2D:1}-\ref{Maxwell 2D:3} and retrieve the space-dependent functions $\varepsilon_r\left(x,y\right)$ and $\mu_r\left(x,y\right)$ defined over the entire domain $\Omega$. Since we are dealing with a full-domain, non-homogeneous retrieval problem, no BCs need to be applied here.
\begin{figure}[h!]
\centering
\includegraphics[width=\linewidth]{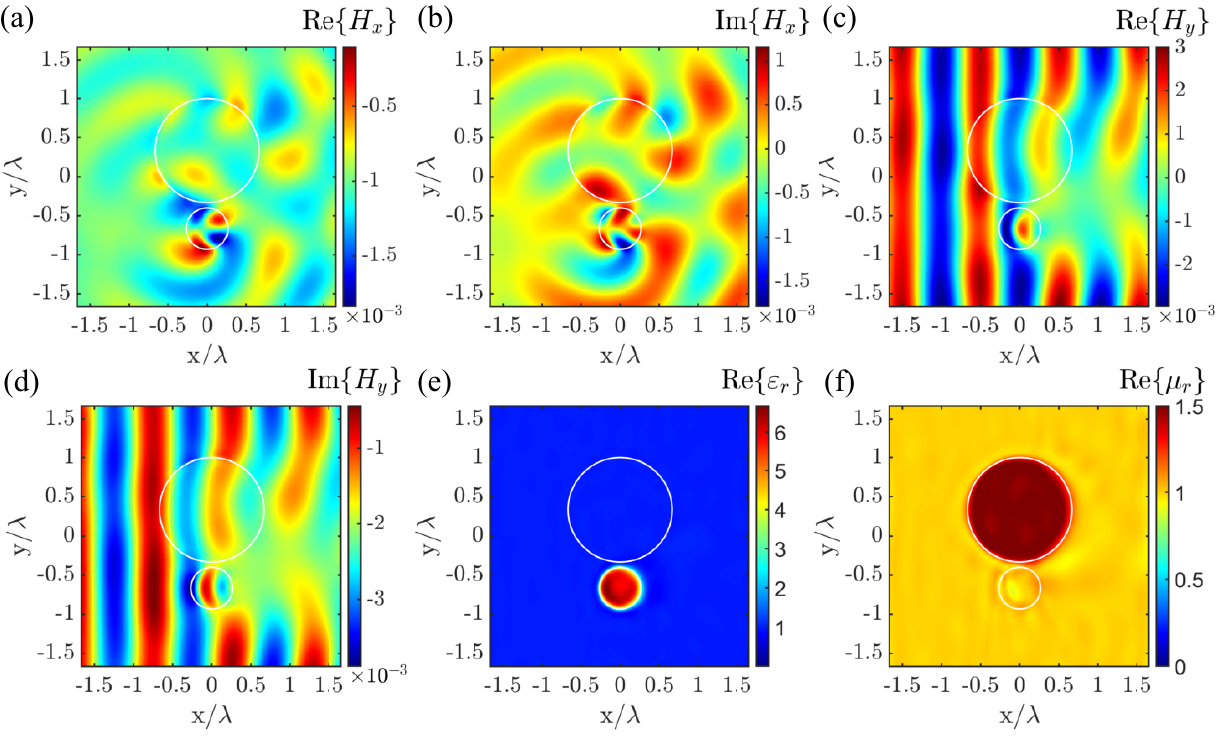}
\caption{(a,b) real and imaginary parts of the complex magnetic field $H_x$ component, respectively.(c,d) real and imaginary parts of the complex magnetic field $H_y$ component, respectively. (e,f) real part of the retrieved complex $\varepsilon_r$ and $\mu_r$ profile by PINNs after $1.5*10^4$ iterations.}
\label{Fig5}
\end{figure}

The investigated dimer has the same dimensions as previously shown in Fig.  \ref{Fig3} (a) except that here we set the optical constants of the two cylinders as $\varepsilon_{r1}=1, \mu_{r1}=1.5+0.5j$ and $\varepsilon_{r2}=6+3j, \mu_{r2}=1$, where one is purely magnetic while the other one is purely dielectric. We run the FEM simulations with settings detailed in the Methods section. The FEM simulation results for the real and imaginary components of $H_x$ used for training PINNs are shown in Fig. \ref{Fig5} (a) and (b), respectively. We display the training datasets $H_y$ real and imaginary parts in Fig. \ref{Fig5} (c) and (d), respectively. The complex $E_z$ field data used for training are shown in the supplementary material. The FCNN parameters and training details are given in the Methods section. We trained the FCNN for $1.5\times{10^{4}}$ iterations before reaching a satisfactory total loss value of $1.5*10^{-2}$. The real part of permittivity and permeability spatial profiles retrieved by PINNs are shown in the Fig. \ref{Fig5} (e) and (f), respectively, which demonstrate accurate reconstruct of each cylinder's shape. We show the retrieved permittivity and permeability imaginary parts in the supplementary material. The obtained complex $\varepsilon_r(x,y)$ and $\mu_r(x,y)$ inside the two cylinder domains have constant values equal to $\varepsilon_{r1}=(1.00\pm0.19)+j(0\pm0.01), \mu_{r1}=(1.48\pm0.07)+j(0.48\pm0.07)$ and $\varepsilon_{r2}=(5.75\pm0.61)+j(3\pm0.36), \mu_{r2}=(1\pm0.04)+j(0\pm0.05)$ , respectively. The maximum $L^2$ error between fields from PINN and FEM simulations is evaluated to be $8\times 10^{-3}$. Therefore, we successfully demonstrated full retrieval of both the particle's shape and values of the electric and magnetic parameters from synthetic electric and magnetic field data. However, in order to obtain stable results with better accuracy at large refractive index contrasts, we need to further generalize the PINNs framework by introducing adaptive weights, as discussed in the next section.

\subsection{Adaptive PINNs: improved accuracy for high-index materials}\label{adpative PINNs}
In all the previous sections we showed that PINNs are suitable for the retrieval of the complex $\varepsilon_r$ and $\mu_r$ of resonant nanostructures from near-field observations outside the objects. However, the solutions of such complex inverse problems become progressively more inaccurate by increasing the refractive index contrast. For instance, we have shown in the supplementary material that the standard PINNs approach loses its accuracy when increasing the object's $\varepsilon_r$ and $\mu_r$ values above a certain threshold value. Therefore, a more accurate and flexible PINN approach needs to be developed where the loss weights in Eq. \ref{eq:loss} are not fixed but can be adaptively modified for the solution of high-index problems. In this section, we in fact demonstrate that additionally training the PINNs' loss weights significantly improves the accuracy of the parameter retrieval for high-index scattering objects. It has been recently demonstrated that adaptive PINNs methods can outperform standard PINNs in accurately solving PDEs with solutions containing sharp transition and and sudden fronts, such as the situations encountered in phase-field PDEs \cite{mcclenny2020self,wight2020solving,wang2020understanding}. The basic idea behind adaptive PINNs is to increase the loss weights for the loss terms that are high. In particular, we apply the following updates for the loss weights at the $k$-th time of $n$ iterations in addition to the standard PINNs:
\begin{eqnarray}
w_f^{k+1} = w_f^{k} + \eta\nabla_{w_f} \mathcal{L}(\boldsymbol{\theta},w_f,w_i,w_b,\varepsilon_{r}) \\
w_i^{k+1} = w_i^{k} + \eta\nabla_{w_i} \mathcal{L}(\boldsymbol{\theta},w_f,w_i,w_b,\varepsilon_{r})  \\
w_b^{k+1} = w_b^{k} + \eta\nabla_{w_b} \mathcal{L} (\boldsymbol{\theta},w_f,w_i,w_b,\varepsilon_{r})
\end{eqnarray}
where $\eta$ is the learning rate for the loss weights. We choose the cylinder with the same geometry as section \ref{section: epsilon single} with $\varepsilon_r=5+1j$ that we have shown in the supplementary material the standard PINNs fail to retrieve. 

\begin{figure}[t!]
\centering
\includegraphics[width=\linewidth]{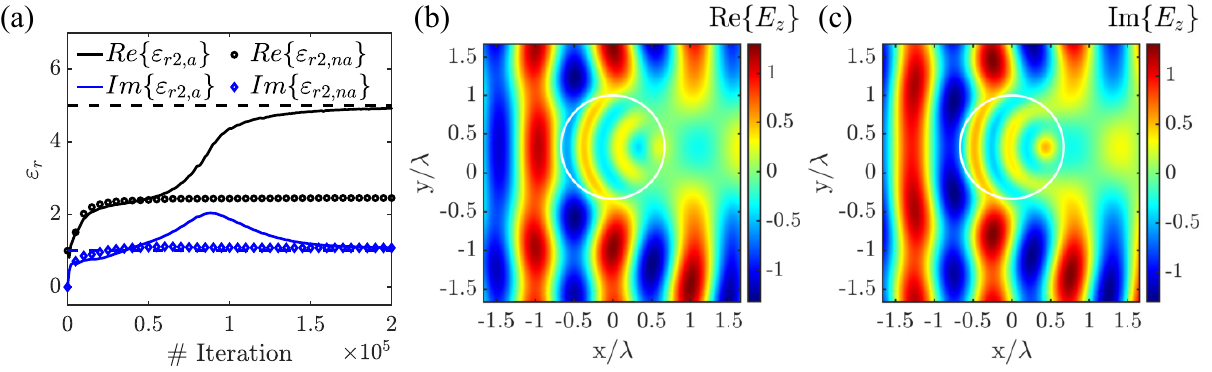}
\caption{(a) complex $\varepsilon_r$ retrieval with respect to iteration number by normal PINN and adaptive PINNs. (b, c) reconstructed real and imaginary $E_z$ profiles obtained by adaptive PINNs, respectively.}
\label{adaptive PINN}
\end{figure}

The complex $\varepsilon_r$ retrieval results with respect to the iteration numbers by using the standard PINNs ($\varepsilon_{r2,na}$) and adaptive PINNs ($\varepsilon_{r2,a}$) are compared in Fig. \ref{adaptive PINN} (a). The same FEM simulation and normal PINN setup as in the section \ref{section: epsilon single} are used. For the adaptive PINNs, we choose $\eta=5$ and update the loss weights by every $5000$ iterations ($n=5000$). We use the fixed loss weight values in standard PINNs as the initial loss weight values for the adaptive PINNs. Further training details are specified in the Methods section. We observe that at the beginning of the training process $\varepsilon_{r2,a}$ and $\varepsilon_{r2,na}$ are close because the initial loss weights for the adaptive PINNs are the same as the fixed loss weights for the standard PINNs. However, as the simulation progresses further, the $\varepsilon_{r2,a}$ converges to its correct value and this value is very different from $\varepsilon_{r2,na}$ at the end of the simulation due to the importance of the loss weight updates. We show the reconstructed complex $E_z$ real and imaginary profiles in Fig. \ref{adaptive PINN} (b) and (c) by using adaptive PINNs, respectively. The $L^2$ errors of the PINNs obtained $E_z$ profiles with respect to FEM simulation are now as low as $1 \times 10^{-4}$ and $2 \times 10^{-4}$ for the real and imaginary parts, respectively. Therefore, the developed adaptive-PINN formulation is suitable for the study of  the complex near-field profile of high-index scatterers and correctly retrieves their complex optical constants in situations where the standard PINNs loses its accuracy entirely. Furthermore, instead of applying the fixed loss weights with values determined by the trial and error procedure, the adaptive PINNs method can balance the interplay between different loss terms automatically. We  demonstrated parameter retrieval for high-index material by using the adaptive PINNs to improve the retrieval accuracy in a 2D configuration. In the last section of our paper, we will introduce the implementation of the general PINN model for complex parameter retrieval of 3D objects with unknown shapes.

\subsection{Complex permittivity retrieval of 3D objects with unknown shapes}\label{section: 3D retrieval}
We finally extend the PINNs framework to retrieve the complex permittivity of 3D objects with unknown shapes and composition, which directly addresses the inverse retrieval of optical parameters of nanostrucutres used in practical biomedical and nanotechnology applications. As a representative example, we show the 3D retrieval of the permittivity profile of non-magnetic objects. However, the presented framework can be modified as shown in the previous sections to additionally retrieve the complex $\varepsilon_r$ and $\mu_r$ 3D profiles simultaneously. By training with the complex electric field data in 3D space and restricting the search based on the wave equation for 3D non-homogeneous media, we demonstrate that PINNs can successfully retrieve the complex permittivity of 3D objects. The implemented non-homogeneous wave equations for non-magnetic objects can be derived from Eq. \ref{nonhomogeneous 2D} as:
\begin{equation}\label{nonhomogeneous 3D}
\nabla^{2} \mathbf{E}-\mu \epsilon \frac{\partial^{2} \mathbf{E}}{\partial t^{2}} -\nabla(\nabla\cdot\mathbf{E})=0
\end{equation}

\begin{figure}[t!]
\centering
\includegraphics[width=\linewidth]{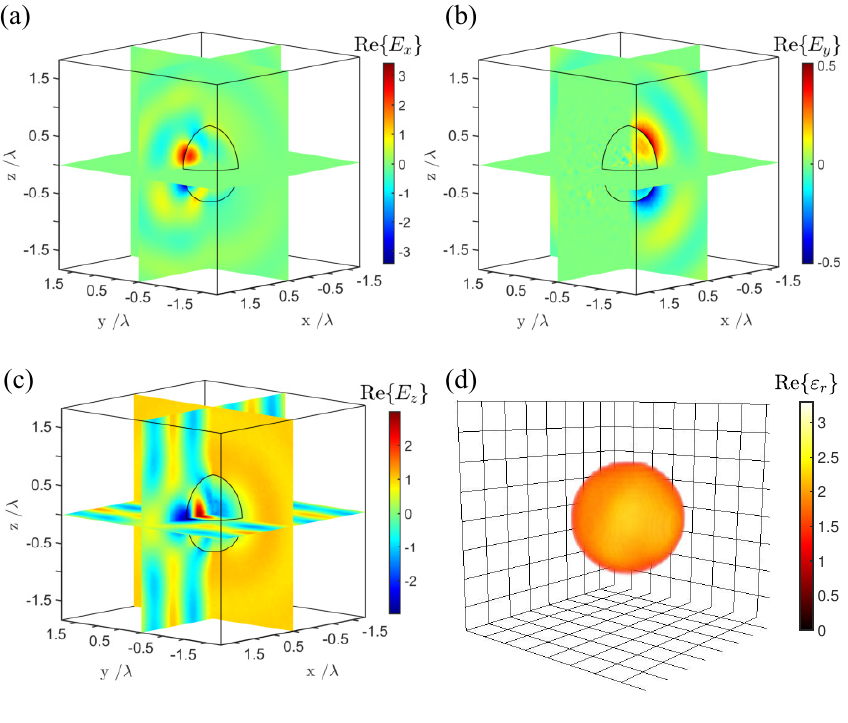}
\caption{(a-c) planar cross sections of the 3D electric field distributions (real part) profiles of $E_x$, $E_y$, and $E_z$ polarization used for training PINN retrieved 3D permittivity profile. (d) The retrieved 3D permittivity profile for a sphere with radius $r=2{\mu}m$ and $\varepsilon=3$. Each grid corresponds to the length equals to 0.5${\mu}m$}
\label{Fig6}
\end{figure}
We consider here a 3D sphere with radius $r_1=2\mu m$ and constant relative permittivity $\varepsilon_{r1}=3$. The complex electric field data $\mathbf{E}=(E_x,E_y,E_z)$  obtained from 3D FEM simulations under plane wave illumination with wavelength $\lambda=3\mu m$ are used as the complex field observations to train PINNs. We display the 3D electric field real part profiles for training PINNs in Fig. \ref{Fig6} (a-c). We employ a FCNN and train it over 120000 iterations until the total loss is below $10^{-1}$. Additional details on the 3D FEM simulations and the network parameters are discussed in our Methods section. The retrieved 3D permittivity is shown in the Fig \ref{Fig6} (d). The obtained non-homogeneous permittivity inside the 3D sphere region is $\epsilon_r=2.57\pm 0.45$, which is in qualitatively good agreement with the ground truth value $\epsilon_1$. This result, which is only limited by our available computational power (we intentionally used a desktop computer for conducting this work as specified in the Methods section), can be further improved by increasing the sampling points for the electric fields in 3D. We conclude by remarking that a similar approach can be applied to extend this PINN framework for the simultaneous 3D retrieval of $\varepsilon_r$ and $\mu_r$ based on 3D near-field microscopy data.
\section{Conclusion}
In conclusion, we have introduced a general DL framework for solving PDEs using PINNs to inversely retrieve unknown 2D and 3D electric and magnetic materials parameters and shape information from synthetic field data. Our results are particularly interesting for inverse microscopy given the current availability of experimental near-field techniques that can measure the optical phase in the near zone with nanoscale resolution. By considering different complex PDE models and field data obtained from FEM simulations, we used PINNs to demonstrate successful retrieval of the complex $\varepsilon_r(x,y)$ and $\mu_r(x,y)$ profiles simultaneously and with very good accuracy. We emphasize that this is achieved within the physics-informed method at significantly reduced data collection and training requirements compared to traditional machine learning approaches that typically employ massive datasets. We presented PINN-based parameter retrieval models that work under both extended and localized excitations that are typically used in SNOM applications. We then proposed and demonstrated an adaptive-PINN algorithm for improving the accuracy of the parameter retrieval for high-index materials. Finally, we showed a successful application of PINNs to the retrieval of the complex permittivity of a 3D scattering object with unknown shape. 
The developed approach can be naturally scaled to any wavelength of interest and applied in arbitrary geometries, providing novel opportunities for non-invasive remote sensing techniques based on measured field data. The proposed computational framework   can naturally enhance existing imaging techniques for the detection of magnetic nanoparticles used in cancer therapy and drugs delivery \cite{Li2016Current,MOHAMMED20171,Cardoso2018Advances} and can be utilized for inspecting and characterizing complex optical devices based on acquired images \cite{Kadic2019}. Although in this paper we were concerned with implementations of PINNs based on complex field data, phase retrieval techniques that recovers the phase information from intensity measurements can be used to solve more general intensity-based (phaseless) retrieval problems with near-field imaging techniques \cite{Gbur2002Tomography,Lei2015phase,Ammari2016phased}.


\section{Methods}
\subsection{FEM simulations}
The complex electric field and magnetic field data are obtained by solving the forward scattering problem using the finite element method \cite{COMSOL}. For the 2D examples, we used a minimum element size of 0.6 nm and perfectly matching layer (PML) boundary with $3\mu$m thickness surrounding the square domain $\Omega$ of side length $10{\mu}m$. The resulting FEM models total degrees of freedom are around $200,000$. We implemented a scattered field formulation and set the background electric field as the plane wave propagating from left to right in the domain $\Omega$. The complex field data are sampled on a $200\times200$ grid points in $\Omega$. \cite{lu2021deepxde}

The same minimum element size and PML boundary thickness are used for the solving 3D forward scattering problem. The degrees of freedom of the 3D FEM model is $\sim 700,000$. The incident plane wave propagates along the x-axis with electric field polarized along $z-axis$. We sampled the complex electric fields on 3D grid with point numbers $50\times50\times30$ along $x-$, $y-$, and $z-$ axis, respectively.

\subsection{Neural network architecture and training hyperparameters}
In all simulations except for the 3D retrieval case, a FCNN with 4 hidden layers and 64 neurons in each hidden layer is trained. We used a FCNN with 3 hidden layers with 20 neurons in each hidden layer for the 3D parameter retrieval. For all the PINNs discussed, we set the learning rate as $10^{-3}$. We fixed $w_i=100$ in the training process for a better convergence to the input field data for the standard PINNs. The adaptive PINNs used $w_f=1$, $w_b=1$, and $w_i=100$ as the initial loss weights. Notice that we use the same hyperparameter values when addressing different problems, which demonstrates the robustness of our methodology for solving parameter retrieval for microscopy. 
We choose the hyperbolic tangent function as the activation function. The Glorot uniform method are used for the ANN weights and biases initialization. The Adam optimizer is used for training the ANN.

The training process is implemented on a desktop with Intel i7-8700K CPU @ 3.70GHz and 32 Gb of RAM, using a Nvidia GeForce GTX 1080Ti GPU. A typical training process for the FCNN takes around 10 h. 

\section*{Data Availability Statement}
The data that support the findings of this study are available from the corresponding author upon reasonable request.

\begin{acknowledgments}
L.D.N. acknowledges the support from the Army Research Laboratory (ARL; Cooperative Agreement Number W911NF-12-2-0023) and the National Science Foundation (ECCS-2015700). The authors would like to thank Dr. Lu Lu and Prof. George Em Karniadakis at Brown University for introducing us to the general methodology of PINNs and for insightful discussions.
\end{acknowledgments}

\bibliography{aipsamp}

\end{document}